\documentclass{article}
\usepackage{amssymb}
\usepackage{amsmath}

\catcode`\@=11

\@addtoreset{equation}{section} \catcode`\@=12 
\DeclareMathOperator{\sgn}{\rm sgn}
\begin{document}

\title{On quantization of the KdV equation}
\author{A.~K.~Pogrebkov \\
Steklov Mathematical Institute, Moscow, Russia\\
pogreb@mi.ras.ru}
\maketitle

\begin{abstract}
Quantization procedure of the Gardner--Zakharov--Faddeev and Magri brackets by means of the
fer\-mi\-o\-nic representation for the KdV field is considered. It is shown that in both cases
the corresponding Hamiltonians are given as sums of two well defined operators. Each of them
is bilinear and diagonal with respect to either fermion, or boson (current)
creation--annihilation operators. As result the quantization procedure needs no any space
cut-off and can be performed on the whole axis. Existence of the solitonic states in the
Hilbert space as well as quantization of soliton parameters are shown to result from this
approach. As a by-product it is also demonstrated that the dispersionless KdV is uniquely and
explicitly solvable in the quantum case.
\end{abstract}

\section{Introduction}

The famous Korteveg--de Vries equation\footnote{In (\ref{KdV}) $\alpha$ is a real parameter
that can be rescaled to 1 when different from 0.}
 \begin{equation}
 v_{t}=6vv_{x}-\alpha _{}^{2}v_{xxx}  \label{KdV}
 \end{equation}
for the real function $v(t,x)$ gave~\cite{1} the first example of the completely integrable
differential equation, i.e.\ equation that can be written in the Lax form as condition of
compatibility of the two linear equations
 \begin{align}
 &-\alpha _{}^{2}\varphi _{xx}^{}(x,k)+v(x)\varphi (x,k)
 =k_{}^{2}\varphi (x,k)  \label{L} \\
 &\varphi _{t}^{}=-4\alpha _{}^{2}\varphi _{xxx}^{}+6v(x)\varphi _{x}^{}+3v_{x}^{}(x)\varphi
 -4i\alpha _{}^{2}k_{}^{3}\varphi \label{M}
 \end{align}
for auxiliary function $\varphi (x,k)$ (we omit $t$-dependence in all cases where it is not
necessary to mention it explicitly). In the case where $v(x)$ is smooth, real function that
decays rapidly enough when $|x|\rightarrow \infty $, the Inverse Spectral Transform (IST)
method (see~\cite{2,3} and citations therein) is applicable to the Eq. (\ref{KdV}). This method
is based on the direct and inverse problems for the spectral problem (\ref{L}) and results in
linear evolution of the corresponding spectral data and unique solvability of the initial
problem for (\ref{KdV}) for all $\alpha \neq 0$.

The KdV equation was also the first nonlinear integrable equation which was proved to be
bi-Ha\-mil\-to\-ni\-an. More exactly, there exist two (local) Poisson structures: the
Gardner--Zakharov--Faddeev~\cite{4} and the Magri~\cite{5} brackets
 \begin{align}
 \{v(x),v(y)\}_{1}^{}& =\delta'(x-y),  \label{GFZ} \\
 \{v(x),v(y)\}_{2}^{}& =(v(x)+v(y))\delta'(x-y)-\frac{\alpha _{}^{2}}{2}\delta'''(x-y),
 \label{Magri}
 \end{align}
such that Eq.~(\ref{KdV}) can be written in any of two forms:
 \begin{equation}
 v_{t}^{}=-\{H_{1}^{},v\}_{1}^{},\qquad v_{t}^{}=-\{H_{2}^{},v\}_{2}^{},  \label{Ham}
 \end{equation}
where Hamiltonians $H_{1}$ and $H_{2}$ are equal to
 \begin{align}
 H_{1}^{}=& \int dx\left( v_{}^{3}(x)+\frac{\alpha _{}^{2}}{2}
 v_{x}^{2}(x)\right) ,  \label{H1} \\
 H_{2}^{}=& \int dx\,v_{}^{2}(x).  \label{H2}
 \end{align}

Quantization of the KdV equation in the periodic case attracted essential attention in the
literature on conformal field theory~\cite{6}--\cite{8} as Poisson structures~(\ref{GFZ})
and~(\ref{Magri}) coincide with commutation relations of current and Virassoro algebras,
respectively. The quantization procedure on the whole axis meets with necessity to supply both
Hamiltonians with proper operator meaning that assumes introduction of some regularization
(say, space cut-off). As the result the any such procedure makes the IST unapplicable already
in the classical case: the continuous and discrete spectra of the problem (\ref{L}) are mixed
and the most interesting, soliton, solutions do not exist.

In this article we show that realization of $v(x)$ as a composite operator in terms of
fermionic fields enables us to avoid necessity of any cut-off procedure for (\ref{H1}) and
(\ref{H2}) as both Hamiltonians are well defined operators in the fermionic Fock space. We
also show that a state that can be called ``1-soliton'' exist in this space.

We already mentioned above that the case of $\alpha=0$ in Eq.~(\ref{KdV}), i.e.\ the case of
dispersionless KdV, needs a special consideration. In this case (see \cite{2}) the IST method
cannot be applied but the initial problem admits a parametric representation of the solution:
$v(t,x)=v_{0}(s)$, $s=x+6tv_{0}(s)$, where $v_{0}(x)$ is the initial data. This representation
describes overturn of the front and in the generic situation $v(t,x)$ is not the uniquely
defined function of $x$ for $t$ large enough. Quantum version of the dispersionless KdV also
attracts essential attention in the literature (e.g.\ \cite{9},~\cite{10}) in the context of
the string theory. Here we show that in the quantum case this equation is uniquely and
explicitly solvable. The article is organized as follows. In the next section we shortly list
some well known results (see, e.g.~\cite{11}) on the ``two dimensional, massless'' fermions.
In the Secs.~3 and 4 we consider quantization of the Gardner--Zakharov--Faddeev~(\ref{GFZ})
and Magri~(\ref{Magri}) brackets, correspondingly. In Sec.~5 we discuss some general aspects
of fermionization of integrable models.

\section{Massless two dimensional fermions}
Let $\mathcal{H}$ denote the fermionic Fock space with vacuum  $\Omega$ generated by operators
$\psi (k)$ and $\psi^{*}(k)$, where $*$ means Hermitian conjugation. These operators obey
canonical anticommutation relations in $\mathcal{H}$,
 \begin{equation}
 \{\psi _{}^{*}(k),\psi (p)\}^{}_{+}=\delta (k-p),\qquad \{\psi (k),\psi (p)\}=0 \label{1}
 \end{equation}
and $\psi (k<0)$ and $\psi ^{*}(k>0)$ are annihilation operators,
 \begin{equation}
 \psi (k)\Omega \Bigr|_{k<0}^{}=0,\qquad \psi _{}^{*}(k)\Omega \Bigr|_{k>0}^{}=0, \label{2}
 \end{equation}
so, correspondingly, $\psi (k>0)$ è $\psi ^{*}(k<0)$ are creation operators. Fermion field is
defined as Fourier transform
 \begin{equation}
 \psi (x)=\frac{1}{\sqrt{2\pi }}\int dk\,e_{}^{ikx}\psi (k)  \label{3}
 \end{equation}
and obeys
 \begin{align}
 \{\psi _{}^{*}(x),\psi (y)\}& =\delta (x-y),\qquad \{\psi (x),\psi (y)\}=0,
 \label{4} \\
 (\Omega ,\psi (x)\psi _{}^{*}(y)\Omega )& =(\Omega ,\psi _{}^{*}(x)\psi (y)\Omega)=
 \frac{(2\pi i)_{}^{-1}}{x-y-i0}  \label{5}
 \end{align}
In description of massless two dimensional Fermi fields  the crucial role is played by two
bilinear combinations, current and density of the energy--momentum tensor:
 \begin{equation}
 j(x)=:\psi _{}^{*}\psi :(x),\qquad u(x)=\frac{i}{2}\left( :\psi _{}^{*}\psi _{x}^{}:(x)-:\psi
 _{x}^{*}\psi :(x)\right).  \label{6}
 \end{equation}
They also will be the main objects of our construction below. In~(\ref{6}) $:\ldots :$ denotes
the Wick ordering with respect to the fermion creation/annihilation operators. In other words,
for example,
 $:\psi ^{*}(x)\psi (y):=\psi ^{*}(x)\psi(y)-(\Omega ,\psi _{}^{*}(x)\psi (y)\Omega )$ and
 $:\psi _{}^{*}\psi :(x)=\lim_{y\rightarrow x}:\psi ^{*}(x)\psi (y):$ and so on. Bilinear
combinations (\ref{6}) are self adjoint operator valued distributions in the space
$\mathcal{H}$ obeying the following commutation relations:
 \begin{align}
 \lbrack \psi (x),j(y)]& =\delta (x-y)\psi (x),  \label{7} \\
 \lbrack j(x),j(y)]& =\frac{i}{2\pi }\delta'(x-y),  \label{8} \\
 \lbrack u(x),u(y)]& =i\left[ (u(x)+u(y))\delta'(x-y)-\frac{1}{24\pi }
 \delta'''(x-y)\right] ,  \label{9} \\
 \lbrack j(x),u(y)]& =ij(y)\delta'(x-y).  \label{900}
 \end{align}
Charge of the Fermi field,
 \begin{equation}
 \Lambda =\int dx\,j(x),  \label{11}
 \end{equation}
is self adjoint operator, it has integer spectrum and annihilates the
vacuum: $\Lambda \Omega =0$.

In what follows we need the following decomposition of operators (\ref{6}):
 \begin{equation}
 j(x)=j_{}^{+}(x)+j_{}^{-}(x),\qquad u(x)=u_{}^{+}(x)+u_{}^{-}(x), \label{12}
 \end{equation}
given by means of projections
 \begin{equation}
 j_{}^{\pm }(x)=\frac{\pm 1}{2\pi i}\int \frac{dy\,j(y)}{y-x\mp i0},\qquad u_{}^{\pm}(x)=
 \frac{\pm 1}{2\pi i}\int \frac{dy\,u(y)}{y-x\mp i0}, \label{13}
 \end{equation}
that admit analytic continuation in the top and bottom half planes of variable $x$. These
positive and negative components are mutually adjoint and
 \begin{equation}
 j^{-}(x)\Omega =0,\qquad u^{-}(x)\Omega =0.  \label{14}
 \end{equation}
Thanks to the first equality and commutation relation~(\ref{8}) bosonic
creation/an\-ni\-hi\-la\-ti\-on operators $j(\pm k)$, $k>0$, can be introduced~\cite{11}:
 \begin{equation}
 j(x)=\frac{1}{2\pi }\int dk\,e_{}^{ikx}j(k),\qquad j_{}^{*}(k)=j(-k),\qquad j(k)\Omega
 |_{k<0}^{}=0.  \label{140}
 \end{equation}
This bosonic operators are bilinear with respect to fermionic ones and they enable introduction
of another Wick ordering for the products of currents, different from the ordering with respect
to the fermionic operators. We denote this normal ordering as $\vdots \ldots \vdots $ that
means that in the expression $\ldots $ positive components of the currents are placed to the
left of the negative ones. Say,
 \begin{equation}
 \vdots j(x)j(y)\vdots =j(x)j(y)-(\Omega ,j(x)j(y)\Omega ),  \label{15}
 \end{equation}
where
 \begin{equation}
 (\Omega ,j(x)j(y)\Omega )=\frac{(2\pi i)_{}^{-2}}{(x-y-i0)_{}^{2}}. \label{16}
 \end{equation}
Then we get
 \begin{equation}
 \vdots j(x)j(y)\vdots
 =j_{}^{+}(x)j_{}^{+}(y)+j_{}^{+}(x)j_{}^{-}(y)+j_{}^{+}(y)j_{}^{-}(x)+j_{}^{-}(x)j_{}^{-}(y)
 \label{17}
 \end{equation}
and again we put $\vdots j^{2}\vdots (x)=\lim_{y\rightarrow x}\vdots j(x)j(y)\vdots$.

Our study of the quantum version of the KdV equation is essentially based on the relations
between these two normal orderings. Say, by~(\ref{15}) and the Wick theorem for the product
$j(x)j(y)$, where currents are given by means of~(\ref{6}) we get
 \begin{equation}
 \vdots j(x)j(y)\vdots =:j(x)j(y):+\frac{1}{2\pi i}\left( \frac{:\psi _{}^{*}(x)\psi (y):-:\psi
 _{}^{*}(y)\psi (x):}{x-y}\right) .  \label{18}
 \end{equation}
In the limit $y\rightarrow x$ we can use that any expression of the type
 $:\ldots \psi(x)\ldots \psi (x)\ldots :$ equals to zero that essentially lowers powers of the
fermionic fields and simplifies results. For example in~(\ref{18}) the first term in the
r.h.s.\ in this limit equals to zero and we get
 \begin{equation}
 \vdots j_{}^{2}\vdots (x)=\frac{1}{\pi }u(x).  \label{19}
 \end{equation}
For the higher powers of currents we also can get relations of this type:
 \begin{equation}
 \vdots j_{}^{3}\vdots (x)=\frac{1}{(2\pi )_{}^{2}}\left( 3:\psi _{x}^{*}\psi
 _{x}^{}:(x)-\frac{1}{2}j_{xx}^{}(x)\right) ,  \label{201}
 \end{equation}
where again the highest powers of Fermi operators cancel out. Eq.~(\ref{18}) can also be used
for the derivatives of currents, say,
 \begin{equation}
 \vdots j_{x}^{2}\vdots (x)=:j_{x}^{2}:(x)+\frac{1}{6\pi }\partial _{x}^{2}u(x)+\frac{1}{6\pi
 i}\left( :\psi _{x}^{*}\psi _{xx}^{}:(x)-:\psi _{xx}^{*}\psi _{x}^{}:(x)\right) , \label{20}
 \end{equation}
where $:j_{x}^{2}:=:\psi _{x}^{*}\psi _{}^{*}\psi _{x}^{}\psi :$, and so on. Bosonic ordering
$\vdots \ldots \vdots $ can be extended for expressions that contain Fermi fields themselves:
 \begin{equation}
 \vdots j(x)\psi (y)\vdots =j_{}^{+}(x)\psi (y)+\psi (y)j_{}^{-}(x). \label{21}
 \end{equation}
Then by (\ref{13})
 \begin{equation}
 \vdots j(x)\psi (y)\vdots =:j(x)\psi (y):+\frac{i}{2\pi }\frac{\psi (x)-\psi(y)}{x-y},
 \label{22}
 \end{equation}
so that this expression as well as its derivatives by $x$ or $y$ are well defined in the limit
$y\rightarrow x$ and we get
 \begin{equation}
 \vdots \frac{\partial ^{n}j}{\partial x_{}^{n}}\psi \vdots (x)=:\frac{\partial ^{n}j}
 {\partial x_{}^{n}}\psi :(x)-\frac{1}{2\pi i(n+1)}\frac{\partial ^{n+1}\psi (x)}
 {\partial x_{}^{n+1}},\quad n=0,1,2,\ldots . \label{23}
 \end{equation}
In particular, for $n=0$ we have the relation
 \begin{equation}
 \vdots j\psi \vdots (x)=-\frac{\psi _{x}^{}(x)}{2\pi i}  \label{24}
 \end{equation}
that results in the bosonization of fermions~\cite{12}--\cite{14}.

\textbf{Remark.} To simplify presentation we put here $\hbar =1$. In order to restore the
Plank constant we need to substitute all commutators and anticommutators $[,]\rightarrow
[,]\hbar $, $\{,\}\rightarrow \{,\}\hbar $ and in the $x$-representation it is also necessary
to substitute $\pi \rightarrow \pi /\hbar $. Taking this into account it is easy to see that
main relations given here are singular in the limit $\hbar \rightarrow 0$, so all of them are
specific for the quantum case and, like fermionization procedure for the
Sine-Gordon~\cite{13},~\cite{14}, have no analog in the classical theory.

\section{Quantization of the Gardner--Zakharov--Faddeev bracket}

In order to quantize the bracket~(\ref{GFZ}) we put
 \begin{equation}
 v(x)=\sqrt{2\pi }j(x),  \label{1-1}
 \end{equation}
so that by~(\ref{8}) $[v(x),v(y)]=i\delta'(x-y)$, and we chose the quantum Hamiltonian as
bosonic ordering of~(\ref{H1}),
 \begin{equation}
 H_{1}^{}=\int dx\vdots v_{}^{3}(x)+\frac{\alpha _{}^{2}}{2} v_{x}^{2}(x)\vdots .  \label{1-2}
 \end{equation}
The main advantage of this approach to quantization of the KdV equation is that this
Hamiltonian is well defined operator as it is, i.e.\ it does not need any regularization.
Indeed, by~(\ref{1-1}) $v^{3}=(2\pi )^{3/2}j^{3}$, so thanks to~(\ref{201})
 \begin{equation}
 H_{1}^{}=\frac{3}{\sqrt{2\pi }}\int dx:\psi _{x}^{*}\psi _{x}^{}:(x)+\alpha _{}^{2}\pi \int
 dx\vdots j_{x}^{2}\vdots (x). \label{1-3}
 \end{equation}
Thus the most singular part of the Hamiltonian~(\ref{1-2}) that was of the third order with
respect to bosonic operators is only of the second order with respect to fermionic operators.
By~(\ref{3}) we have that
 \begin{equation}
 \int dx:\psi _{x}^{*}\psi _{x}^{}:(x)=\int\limits_{-\infty }^{0}dk\,k_{}^{2}\psi_{}^{*}(k)
 \psi (k)-\int\limits_{0}^{\infty }dk\,k_{}^{2}\psi (k)\psi _{}^{*}(k), \label{1-4}
 \end{equation}
while by~(\ref{140})
 \begin{equation}
 \int dx\vdots j_{x}^{2}\vdots (x)=2\int\limits_{0}^{\infty }dk\,k_{}^{2}j(k)j(-k).
 \label{1-5}
 \end{equation}
Thus we see that both terms in~(\ref{1-3}) are bilinear with respect to the fermionic, or
bosonic creation--annihilation operators, they are normally ordered and have a diagonal form,
i.e.\ they include ``creation$\times$annihilation'' terms only. In particular, by~(\ref{3})
and~(\ref{140})
 \begin{equation}
 H_{1}^{}\Omega =0,  \label{1-6}
 \end{equation}
while the first term in~(\ref{1-3}) is not positively defined as follows from (\ref{1-4}). On
the other side~(\ref{1-6}) demonstrates that vacuum expectation values for equal time products
of fields do not depend on time and then both types of Wick ordering are compatible with time
evolution. Thanks to~(\ref{20}) we can rewrite the Hamiltonian in terms of the fermionic
ordering as follows:
 \begin{align}
 H_{1}^{}& =\frac{3}{\sqrt{2\pi }}\int dx:\psi _{x}^{*}\psi_{x}^{}:(x)+
 \alpha _{}^{2}\pi \int dx:j_{x}^{2}:(x)+  \nonumber \\
 & +\frac{\alpha _{}^{2}}{6i}\int dx\left( :\psi _{x}^{*}\psi _{xx}^{}:(x)-
 :\psi _{xx}^{*}\psi_{x}^{}:(x)\right) .  \label{1-7}
 \end{align}

It is clear that time evolution given by the Hamiltonian~(\ref{1-2})
 \begin{equation}
 v_{t}^{}\equiv i[H_{1}^{},v]=6\vdots vv_{x}^{}\vdots -\alpha _{}^{2}v_{xxx}^{} \label{1-8}
 \end{equation}
is exactly the quantum version of the Eq.~(\ref{KdV}), normally ordered with respect to the
bosonic operators. Thanks to~(\ref{1-1}) in terms of the current Eq.~(\ref{1-8}) reads as
$j_{t}^{}(x)=(3\sqrt{2\pi }\vdots j^{2}\vdots (x)-\alpha ^{2}j_{xx}(x))_{x}$ that in
correspondence with~(\ref{1-3}) gives the quantum \textbf{bilinear} form of the KdV equation
in terms of Fermi field:
 \begin{equation}
 j_{t}^{}(x)=\frac{\partial }{\partial x}\left( \frac{6}{\sqrt{2\pi }} u(x)-\alpha_{}^{2}
 j_{xx}^{}(x)\right) ,  \label{1-9}
 \end{equation}
where $u$ is defined in~(\ref{6}). Eq.~(\ref{1-9}) can be considered as ``quantum Hirota form''
of the KdV equation. On the other side time evolution $\psi _{t}\equiv i[H_{1},\psi ]$ of the
Fermi field is nonlinear,
 \begin{align}
 \psi _{t}^{}(x)& =\frac{3i}{\sqrt{2\pi }}\psi _{xx}^{}(x)+2\pi i\alpha_{}^{2}
 \vdots j_{xx}^{}\psi \vdots (x)=  \nonumber \\
 & =\frac{3i}{\sqrt{2\pi }}\psi _{xx}^{}(x)-\frac{\alpha _{}^{2}}{3}\psi _{xxx}^{}(x)+
 2\pi i\alpha _{}^{2}:j_{xx}^{}\psi :(x),  \label{1-10}
 \end{align}
where the first line follows from~(\ref{1-3}) and second line from~(\ref{1-7}) (or
from~(\ref{23}) for $n=2$).

Study of the spectrum of the quantum Hamiltonian must be subject for the special investigation.
Here we demonstrate only that in the fermionic Fock space there exists state that can be
considered as 1-soliton in the sense that average of the field $v$ with respect to this state
equals to the famous classical 1-soliton solution
 \begin{equation}
 v_{\text{1-solit}}^{}=-\frac{2\kappa ^{2}}{\displaystyle\cosh ^{2}\frac{\kappa }{\alpha }
 (x-4\kappa _{}^{2}t-\alpha )},  \label{1sol}
 \end{equation}
at least at zero (or any fixed) value of time. For this sake consider unitary transformation
 \begin{equation}
 \widehat{\psi }(x)=W\psi (x)W_{}^{*},\qquad \widehat{j}(x)=Wj(x)W_{}^{*}, \label{1-11}
 \end{equation}
such that
 \begin{equation}
 \widehat{\psi }(x)=e^{i\gamma (x)}\psi (x),  \label{hat}
 \end{equation}
where $\gamma (x)$ is some (smooth) $c$-valued function. In order that operator $W$ exists in
$\mathcal{H}$ it is necessary that this function obeys asymptotic condition
 \begin{equation}
 \lim_{x\rightarrow +\infty }\gamma (x)-\lim_{x\rightarrow -\infty }\gamma(x)=
 2\pi n,\quad n\in \mathbb{Z}.  \label{cond:g}
 \end{equation}
Then for the current we get
 \begin{equation}
 \widehat{j}(x)=j(x)-\frac{\gamma'(x)}{2\pi },  \label{hat:j}
 \end{equation}
so that by~(\ref{1-1}) we get
 $(W^{*}\Omega ,v(x)W^{*}\Omega )=-\gamma'(x)/(2\pi )$. In order
to equal the r.h.s.\ to the value of solution~(\ref{1sol}) for $t=0$ we have to put
 \begin{equation}
 \gamma (x)=2\alpha \kappa \sqrt{2\pi }\tanh \frac{\kappa }{\alpha }(x-x_{0}^{}).  \label{g1}
 \end{equation}
Then by~(\ref{cond:g}) we get that the soliton variable $\kappa $ is quantized:
 \begin{equation}
 \kappa =\frac{\sqrt{2\pi }}{4\alpha }n,\quad n=0,1,2,...  \label{cond:k1}
 \end{equation}
Let us mention that $n=0$ in fact corresponds to the case where
 $(W^{*}\Omega ,v(x)W^{*}\Omega)=0$, i.e.\ soliton is absent. This is the only value that
preserves the charge $\Lambda $~(\ref{11}) under transformation~(\ref{1-11}),~(\ref{hat}). In
the standard quantization of the KdV equation the bosonic Hilbert space corresponds to the
zero charge sector of the fermionic Fock space ${\cal H}$ here. In this sector
transformation~(\ref{hat:j}) where $n$ in~(\ref{cond:k1}) is positive does not exist. This
proves that the soliton state $W^{*}\Omega$ that appeared in our construction there exists
only thanks to the fermionization procedure.

We proved in~(\ref{1-9}) that with time the current (and thus $v(t,x)$) evolves as bilinear
combination. Thus decomposition in positive and negative parts given in~(\ref{12}) is
preserved under evolution and these parts of $j$ and $u$ evolve independently. This proves
correctness of the following construction. Let $\varphi (x,k)$, where $k\in\mathbb{C}$, be
operator solution of the differential equation
 \begin{equation}
 -\alpha _{}^{2}\varphi _{xx}^{}+\vdots v(x)\varphi \vdots =k_{}^{2}\varphi ,  \label{1-12}
 \end{equation}
where we understand $\vdots \ldots \vdots $ as
 \begin{equation}
 \vdots v(x)\varphi \vdots =v_{}^{+}(x)\varphi +\varphi v_{}^{-}(x) \label{1-13}
 \end{equation}
and we choose solution of~(\ref{1-12}) that is normalized at $k$-infinity as
 $\lim_{k\rightarrow \infty }\varphi (x,k)e^{ikx/\alpha }=1$. This differential equation
is nothing but explicit quantization of the classical spectral problem~(\ref{L}). Solution (at
least formal) of~(\ref{1-12}) exists and equals to the ordered (in the bosonic sense) classical
Jost solution with substitution of the $v(x)$ for its operator expression. Then taking
independence of the normal ordering on time into account we derive that $\varphi$ obeys the
quantized version of~(\ref{M}),
 \begin{equation}
 \varphi _{t}^{}=-4\alpha _{}^{2}\varphi _{xxx}^{}+6\vdots v(x)\varphi _{x}^{}\vdots +3\vdots
 v_{x}^{}(x)\varphi \vdots -4i\alpha _{}^{2}k_{}^{3}\varphi .  \label{1-14}
 \end{equation}
Now, by~(\ref{1-13}) and again by independence of the normal ordering on time we get that
difference $\varphi_t-i[H_1,\varphi]$ obeys~(\ref{1-2}). Then thanks to the asymptotic
condition on $\varphi$ this difference equals to zero, i.e.\ $\varphi_t=i[H_1,\varphi]$. On
the other side in analogy with the classical case condition of compatibility of
Eqs.~(\ref{1-12}) and~(\ref{1-14}) is equivalent to
 \begin{equation}
 \vdots \left( v_{t}^{}-6vv_{x}^{}+\alpha _{}^{2}v_{xxx}^{}\right) \varphi (x,k)\vdots =0,
 \label{1-15}
 \end{equation}
that gives quantum KdV equation~(\ref{1-8}) thanks to the asymptotic property of
$\varphi(x,k)$.

As we mentioned in Introduction the dispersionless KdV needs special consideration in the
classical case. Consider now this case of $\alpha =0$ in the quantum situation. First of all,
in this case only the first term (bilinear with respect to fermion operators) in the
Hamiltonian $H_{1}$ in~(\ref{1-3}) is preserved. Correspondingly, by~(\ref{1-10}) the Fermi
field evolves linearly in this case and we get explicit solution
 $\psi (t,x)=(2\pi )^{-1}\int dy\int dk\,\exp \{ik(x-y)-3ik^{2}t/\sqrt{(2\pi )}\}\psi (y)$,
where $\psi (x)$ is the initial value of $\psi (t,x)$. Then
 \begin{equation}
 \psi (t,x)=\frac{e_{}^{-\frac{i\pi }{4}\sgn t}}{\sqrt{6|t|\sqrt{2\pi }}} \int
 dy\,e_{}^{\frac{i(x-y)^{2}}{12t}\sqrt{2\pi }}\psi (y).  \label{1-16}
 \end{equation}
Thanks to this relation it is easy to check directly that equal time expectation values of
products of the Fermi fields are time-independent, say,
 $(\Omega ,\psi (t,x)\psi ^{*}(t,y)\Omega )=(\Omega,\psi (x)\psi ^{*}(y)\Omega )$ that
confirms that normal ordering is preserved under time evolution. Thus by~(\ref{6})
and~(\ref{1-1}) we get explicit operator solution
 \begin{equation}
 v(t,x)=\frac{1}{6|t|}\int dy\int dy'\,e_{}^{\frac{\sqrt{2\pi }}
 {12it}((x-y)^{2}-(x-y')^{2})}:\psi _{}^{*}(y)\psi (y'): \label{1-17}
 \end{equation}
of the quantum dispersionless KdV
 \begin{equation}
 v_{t}^{}(t,x)=6\vdots v(t,x)v_{x}^{}(t,x)\vdots.   \label{1-18}
 \end{equation}
In fact it is not complicated to check directly that~(\ref{1-17}) obeys this equation. Thus we
see that in the quantum case dispersionless equation is uniquely and explicitly solvable.

\section{Quantization of the Magri bracket}

In this section we describe in short some rezults on quantization of the Magri
bracket~(\ref{Magri}). Fermionization procedure of the corresponding commutation relation is
well known in the literature (see, e.g.,~\cite{7}), so here we concentrate mainly on the
structure of the Hamiltonian that results from this procedure. Taking
Eqs.~(\ref{8})--(\ref{900}) into account it is reasonable to choose
 \begin{equation}
 v(x)=u(x)+\alpha \sqrt{\pi }j_{x}^{}(x),  \label{2-1}
 \end{equation}
where $\alpha $ at the moment is an arbitrary real constant. Thanks to~(\ref{6}) and
(\ref{19}) this operator can also be written in the following forms
 \begin{equation}
 v(x)=\left( \frac{i}{2}+\alpha \sqrt{\pi }\right) :\psi _{}^{*}\psi_{x}^{}:(x)+
 \left( \frac{-i}{2}+\alpha \sqrt{\pi }\right) :\psi_{x}^{*}\psi :(x)\equiv
 \pi \vdots j_{}^{2}\vdots (x)+\alpha \sqrt{\pi } j_{x}^{}(x).
 \label{2-2}
 \end{equation}
Thanks to~(\ref{8})--(\ref{900}) it obeys the commutation relations
 \begin{equation}
 \lbrack v(x),v(y)]=i\left[ (v(x)+v(y))\delta'(x-y)-\left( \frac{\alpha _{}^{2}}{2}+
 \frac{1}{24\pi }\right) \delta'''(x-y)\right] ,
 \label{2-3}
 \end{equation}
that is natural quantum analog of the bracket~(\ref{Magri}) and involves also quantum
corrections (see concluding Remark of Sec.~2). In order to introduce a quantum version of the
Hamiltonian~(\ref{H2}) we need some adequate notion of ordering for operators~(\ref{2-1}). We
choose here
 \begin{equation}
 {\text{N}}(v_{}^{2})(x)=v_{}^{+}(x)_{}^{2}+2v_{}^{+}(x)v_{}^{-}(x)+ v_{}^{-}(x)_{}^{2},
 \label{2-4}
 \end{equation}
where positive and negative parts are understood in the sense of~(\ref{13}). Let us mention,
that now, contrary to bosonic Wick ordering $\vdots\ldots\vdots$, the difference of $v^{2}(x)$
and ${\text{N}}(v^{2})(x)$ is some (infinite) operator. Generally speaking this means that
also some other orderings must be considered and with this respect our results below have to
be considered as preliminary. Let
 \begin{equation}
 H_{2}^{}=\int dx\,{\text{N}}(v_{}^{2})(x),  \label{2-5}
 \end{equation}
then
 \begin{equation}
 v_{t}^{}(x)\equiv i[H_{2}^{},v(x)]=\frac{\partial }{\partial x}\left[3{\text{N}}
 (v_{}^{2})(x)- \alpha _{}^{2}v_{xx}(x)\right] .  \label{2-6}
 \end{equation}
We see that parameter $\alpha $ introduced in~(\ref{2-1}) can be identified with the parameter
$\alpha$ in the KdV equation. Moreover, by the calculations of the same type like in previous
section we derive that the Hamiltonian is equal to
 \begin{equation}
 H_{2}^{}=\left( \frac{1}{4}+\alpha _{}^{2}\pi \right) \int dx\vdots j_{x}^{2}\vdots
 (x)-i\alpha _{}^{2}\int dx\left( :\psi _{x}^{*}\psi _{xx}^{}:(x)-:\psi _{xx}^{*}
 \psi_{x}^{}:(x)\right) .  \label{2-7}
 \end{equation}
Again, this operator is well defined and equals to the sum of two terms bilinear with respect
to bosonic and fermionic creation/annihilation operators, correspondingly. In particular, for
the dispersionless case
 \begin{equation}
 H_{2}^{}\Bigr|_{\alpha =0}^{}=\frac{1}{4}\int dx\vdots j_{x}^{2}\vdots(x),  \label{2-8}
 \end{equation}
so that now the Hamiltonian is bilinear with respect to bosonic operators. Again this means
that in this case the dispersionless KdV is explicitly solvable. Let us mention that the case
$\alpha =0$ considered here has nothing in common with the known from the literature so called
case $c=-2$, see for example~\cite{8}, where observation made in~(\ref{2-8}) was missed.

\section{Conclusion}

We demonstrated here that under fermionization procedure where the KdV field is given as
composite object the highest operator terms of the both Hamiltonians become just bilinear with
respect to the Fermi fields. Moreover, both Hamiltonians are well defined operators and need
no any regularization, so that the quantum theory can be considered on the whole axis. The
fact that under fermionization the nonlinear terms of the Hamiltonians are essentially
simplified is known from the first works on this procedure for the Sine-Gordon
equation~\cite{12}--\cite{14}. Fermi fields also appear naturally in the quantum theory of the
Nonstationary Schr\"{o}dinger equation and some integrable models of the Statistical physics
(see~\cite{15}--\cite{17}) as describing the limit of the infinite constant.

Surprisingly, but the ``fermionic" properties of the integrable models can be observed already
on the classical level. Indeed, if $N$-soliton solutions of the KdV with parameters
$\kappa_{1},\ldots ,\kappa _{N}$ (cf.\~(\ref{1sol})) are considered then the IST
method~\cite{2},~\cite{3} proves that these parameters never coincide,
 \begin{equation}
 \kappa _{i}^{}\neq \kappa _{j}^{},\qquad i\neq j.  \label{kappa}
 \end{equation}
In other words, in spite of the fact that $\kappa_{1},\ldots ,\kappa _{N}$ are action
variables~\cite{2}, the corresponding classical phase space is not flat. In the quantum case
condition ~(\ref{kappa}) is a kind of Pauli principle that is characteristic for fermions.

We proved existence of the solitonic state and quantization~(\ref{cond:k1}) of the soliton
parameter but the problem of investigation of the spectra of both ``bilinear" Hamiltonians is
left open and needs a special consideration. Let us mention also that results obtained here
are based on the properties of the standard massless fermion fields. It is interesting to
consider the analogous procedure based on (anyonic) generalization of this object~\cite{18}.

This work is supported in part by Russian Foundation for Basic Research (grants \# 99-01-00151
and 00-15-96046) and by INTAS (grant \# 99-1782).


\begin{thebibliography}{99}

\bibitem{1}  C.~S.~Gardner, J.~M.~Green, M.~D.~Kruskal, and R.~M.~Miura
\textsl{Phys. Rev. Lett.} \textbf{19} (1967) 1095

\bibitem{2}  S.~P.~Novikov, S.~V.~Manakov, L.~P.~Pitaevsky, and V.~E.~Zakharov
\textsl{Theory of Solitons. The Inverse Scattering Method}, New York, (1984)

\bibitem{3}  F.~Calogero and A.~Degasperis \textsl{Spectral Transform and Solitons}, vol. 1,
Amsterdam, North-Holland (1982)

\bibitem{4}  C.~Gardner \textsl{Journ. Math. Phys.} \textbf{12} (1971) 1548;
V.~E.~Zakharov, L.~D.~Faddeev, \textsl{Funct. Anal. Appl.} \textbf{5} (1971) 18 (Russian
pages)

\bibitem{5}  F.~Magri \emph{``A geometrical approach to the nonlinear solvable equations''} in
\textsl{Nonlinear Evolution Equations and Dynamical Systems} (Lectures Notes in Physics,
vol.~120), eds. M.~Boiti, F.~Pempinelli, and G.~Soliani, Springer, Berlin, p.~233 (1980)

\bibitem{6} P.~B.~Wiegmann, A.~Zabrodin \textsl{Commun. Math.
Phys.} \textbf{213} (2000) 523

\bibitem{7}  V.~V.~Bazhanov, S.~L.~Lukyanov, A.~B.~Zamolodchikov \textsl{Commun. Math. Phys.}
\textbf{177} (1996) 381

\bibitem{8}  P.~Di~Francesco, P.~Mathieu, D.~S\'{e}n\'{e}chal \textsl{Mod Phys.
Lett.} \textbf{A7} (1992) 701

\bibitem{9}  J.~Barcelos-Neto, A.~Constandache, A.~Das \textsl{Phys.Lett. A} \textbf{268} (2000)
342

\bibitem{10}  I.~Krichever \textsl{Commun. Math. Phys.} \textbf{143} (1992) 415--429;
\textsl{Commun. Pure Appl. Math.} \textbf{47} (1992) 437--476

\bibitem{11}  A.~Wightman \textsl{Intruduction to Some Aspects of the Relativistic
Dynamics of Quantized Fields}, Princeton--New Jersey, Princeton Univ. Press (1964)

\bibitem{12}  S.~Coleman \textsl{Phys. Rev.} \textbf{D11} (1975) 2088

\bibitem{13}  S.~Mandelstam \textsl{Phys. Rev.} \textbf{D11} (1975) 3026

\bibitem{14}  A.~K.~Pogrebkov, V.~N.~Sushko \textsl{Theor. Math. Phys.}
\textbf{24} (1975) 937; \textbf{26} (1976) 286

\bibitem{15}  F.~Colomo, A.~G.~Izergin, V.~E.~Korepin, and V.~Tognetti
\textsl{Theor. Math. Phys.} \textbf{94} (1993) 11

\bibitem{16}  N.~M.~Bogoliubov, A.~G.~Izergin, and V.~E.~Korepin \textsl{Quantum
Inverse Scattering Method and Correlation Functions}, Cambridge, Cambridge Univ. Press (1989)

\bibitem{17}  N.~A.~Slavnov \textsl{Theor. Math. Phys.} \textbf{108} (1996)
993

\bibitem{18}  N.~Ilieva and W.~Thirring \textsl{Eur. Phys. Journ.} \textbf{C6}
(1999) 705; \textsl{Theor. Math. Phys.} \textbf{121} (1999) 1294

\end{thebibliography}
\end{document}